# Effect of Robo-taxi User Experience on User Acceptance: Field Test Data Analysis

Sunghee Lee[1], Soyoung Yoo[1], Seongsin Kim[1], Eunji Kim[1], Namwoo Kang[2,*]

[1]Department of Mechanical Systems Engineering,
Sookmyung Women's University, 04310, Seoul, Korea

[2]The Cho Chun Shik Graduate School of Green Transportation,
Korea Advanced Institute of Science and Technology, 34141 Daejeon, Korea

[*]Corresponding author: nwkang@kaist.ac.kr

**Abstract**

With the advancement of self-driving technology, the commercialization of Robo-taxi services is just a matter of time. However, there is some skepticism regarding whether such taxi services will be successfully accepted by real customers due to perceived safety-related concerns; therefore, studies focused on user experience have become more crucial. Although many studies statistically analyze user experience data obtained by surveying individuals' perceptions of Robo-taxis or indirectly through simulators, there is a lack of research that statistically analyzes data obtained directly from actual Robo-taxi service experiences. Accordingly, based on the user experience data obtained by implementing a Robo-taxi service in the downtown of Seoul and Daejeon in South Korea, this study quantitatively analyzes the effect of user experience on user acceptance through structural equation modeling and path analysis. We also obtained balanced and highly valid insights by reanalyzing meaningful relationships obtained through statistical models based on in-depth interview results. The results revealed that the experience of the traveling stage had the greatest effect on user acceptance, and the cutting edge of the service and apprehension of technology were emotions that had a significant effect on user acceptance. Based on these findings, we suggest guidelines for the design and marketing of future robot taxi services.

# 1. Introduction

With the advent of the sharing economy and Robo-taxis (or self-driving taxis), the automotive industry is facing new technological, social, and regulatory changes. The Robo-taxi is expected to alleviate traffic congestion and reduce the need for parking through an active car-sharing service, as well as lower carbon emissions through optimized operation by sharing connected road information (1). Furthermore, it will be a low-cost, affordable, and easily accessible option for people in the outskirts of cities or rural areas where advanced public transportation is not available (2). In this changing environment, multiple automakers, IT firms, and shared service companies are quickly seizing the initiative in the Robo-taxi market through the establishment of various partnerships (3–8).

However, high-quality technology is not always necessarily accepted by consumers. Since self-driving is a technology directly related to safety, relieving the user's anxiety is a significant hurdle. As a result, various user-centered studies related to Robo-taxis have been conducted. Many studies have used a survey approach to analyze the relationship between the user and self-driving vehicles or the Robo-taxi (9–11), while others have conducted experiments on the interaction between the taxi and the user in an indirect way using simulation and VR (12–16). Recently, studies that test the interaction between the user and the Robo-taxi based on actual field tests have also been conducted to overcome the limitations of virtual experiments conducted through surveys and simulators (17–22). User response can be analyzed using quantitative statistical methods as well as qualitative interviews. For example, some studies analyze factors affecting the acceptance of self-driving technology through structural equation modeling (SEM) or path analysis based on customer surveys (23–31).

This study performs SEM and path analysis based on actual user experience data for the Robo-taxi service to analyze the factors affecting user acceptance. The differences between previous studies and this study are as follows. First, previous studies on SEM used survey data for people who had no experience with the Robo-taxi service. However, responses based on imagination about new technologies and services that they have never experienced before would differ greatly from reality. Specifically, regarding services directly related to safety, there are significant differences in user response before and after the experience (22). This study differs in that it collected survey data after having people experience an unmanned taxi service. Second, we compared the in-depth interview results and simultaneously analyzed videos to interpret the meaning of the statistical model analysis results. By performing quantitative and qualitative analyses in a balanced manner, we obtained more explainable and valid insights.

We constructed and analyzed three models using the user experience data on the Robo-taxi obtained from previous studies (21–22) performed by the authors. We analyzed the first model using path analysis to identify how the main evaluation factors for the quality of the Robo-taxi service by stage (i.e., call, pick-up, traveling, drop-off) affect Robo-taxi service satisfaction. We analyzed the second model using SEM to identify how the Robo-taxi service experience affects user acceptance. We analyzed the third model with SEM to identify how positive and negative emotions towards the Robo-taxi service experience affect user acceptance.

The remainder of this paper is organized as follows. Section 2 reviews the related previous works. Section 3 introduces the hypotheses and theoretical background of this study. Section 4 presents the field test process for obtaining user experience data. Section 5 introduces three models for hypothesis testing and describes the analysis results. Section 6 analyzes the model results in detail and derives the implications. Section 7 summarizes the study and describes its limitations.

# 2. Related Works

Our target research area is SEM studies using the Robo-taxi field test data, and related works are introduced in this section. First, in Section 2.1, research on user-centered self-driving technology is classified into three categories in terms of data collection methods: simple surveys, simulators, and field tests. Second, in Section 2.2, we introduce studies on SEM for self-driving experience. Lastly, Section 2.3 describes a research gap based on the limitations of previous studies.

## 2.1. User-centered self-driving technology research

### 2.1.1. Survey-based user research

Hohenberger et al. (9) surveyed 1,603 people and found that people with high self-enhancement had low anxiety about self-driving vehicles. Hulse et al. (10) surveyed 1,000 people in the UK on road safety and user acceptance. They argued that self-driving vehicles would be more attractive to male and young participants. Tussyadiah et al. (11) surveyed 325 people on the influence of attitude and trust in technology on the intention to use self-driving taxis. They argued that consumers should trust that the Robo-taxi would work as designed and that expectations of reliability, functionality, and usefulness contribute to their intention to use.

### 2.1.2. Simulator-based user research

Studies have used simulation and VR to investigate the relationship between Robo-taxis and the user. Koo et al. (12) conducted experiments with 64 participants using a driving simulator equipped with automatic braking and presented an interaction model for the user to communicate with self-driving vehicles. The driving performance felt by drivers was improved when providing "reason" information as to why the vehicle behaved as it did rather than when providing the vehicle's "behavior" information. Koo et al. (13) conducted simulation experiments with 40 participants and showed that appropriate voice alerts alleviated driver anxiety. Cho et al. (14) tested 68 participants using a driving simulator with different automation levels and showed that the anxiety was highest for automation level 3 and dropped slightly for automation level 4. Jamson et al. (15) conducted experiments on 49 participants using a driverless car simulator and showed that drivers were willing to give up supervisory responsibilities. Griesche et al. (16) examined the relationship of preference between the driver's driving method and that of self-driving vehicles. The results revealed that most drivers liked a driving style similar to their own, and all participants did not like small safety margins and high acceleration driving styles.

### 2.1.3 Field-test based user research

Recently, user studies through field tests have been conducted to overcome the limitations of surveys and simulators. Rothenbucher et al. (17) modified the driver's seat of a vehicle in such a way that pedestrians could not see the driver; thus, it was recognized as a fully autonomous vehicle. They investigated pedestrians' responses and experiences with self-driving vehicles on a university campus. Kim et al. (18) implemented a taxi service within a university campus using a real self-driving vehicle and tested passengers' responses and the validity of self-driving driving technology. Banks et al. (19) collected video data on user behavior through a field test of a partially automated self-driving vehicle and analyzed the relationship between self-driving function and user behavior. Zoellick et al. (20) tested user acceptance after experiencing a shared autonomous electric vehicle in Berlin. Kim et al. (18) tested user experience in Daejeon in Korea by implementing the Robo-taxi service. They suggested a solution that could compensate for the shortcomings of self-driving technology by introducing the concept of a virtual stop. Yoo et al. (22) analyzed the anxiety factors of Robo-taxis based on a field test. They designed and implemented a Robo-taxi human-machine interaction (HMI) that could relieve anxiety and tested the service in the downtown area of Seoul, Korea.

## 2.2. SEM-based self-driving technology research

Regarding studies on self-driving vehicles and the Robo-taxi, Payre et al. (23) analyzed technology acceptability, attitudes, personality characteristics, and the intention to use self-driving vehicles with SEM. They surveyed 153 male drivers and observed a strong positive correlation between attitudes and the intention to use self-driving vehicles. Rödel et al. (24) surveyed 336 people and investigated the factors of user acceptance and user experiences, such as the ease of use for self-driving vehicles, attitudes toward system use, cognitive behavior control, and behavioral intention. Ro and Ha (25) surveyed 1,506 participants and showed that convenience, safety, ethics, licenses, and cost had a direct effect on the acceptance attitude toward self-driving vehicles, while convenience, safety, and financial cost had a direct effect on the intention to use self-driving vehicles. Bennett et al. (26) conducted a survey of 211 blind people, showing that hope for independence, concern over safety, and affordability affect the acceptance of autonomous vehicles. Rahimi et al. (27) presented an analysis of user acceptance for shared mobility and autonomous vehicles based on user characteristics (e.g., age, education, income, etc.). Zhu et al. (28) found that mass media and social media had different impacts on the acceptance of self-driving cars, based on a survey of 355 college students. Wu et al. (29) showed that green perceived usability, perceived ease of use, and environmental concerns have a positive relationship with the user acceptance of autonomous EVs. Acheampong and Cugurullo (30) analyzed user acceptance of general self-driving cars, self-driving car-sharing services, autonomous public transport services, and self-driving car ownership, respectively. Manfreda (31) tested millennials for smart cities, and the results showed that the perceived benefits of AVs affect AV adoption, and perceived safety can reduce concerns about AVs.

### 2.3. Research gap

To the best of our knowledge, no SEM research has been conducted based on actual user experience data obtained through the Robo-taxi field test. Robo-taxis have significant differences in terms of real-life experience and imagination. In particular, it is difficult to analyze the perception of the safety of Robo-taxis with imagination without experience. Therefore, the results of the SEM studies without field testing are limited in terms of reliability. In this study, we use actual user experience data obtained through field tests to propose user acceptance SEM and emotion SEM. In addition, quantitative analysis results were complemented through qualitative analysis through in-depth interviews.

## 3. Hypothesis

### 3.1. Effect of user experience on user acceptance

User perception of new technology affects user acceptance (32). Users who do not have direct experience in a given area perceive new technology by judging it on an abstract basis but could judge based on more specific criteria after direct experience (33). Because users judge based on experience, user experience has been heavily covered in multiple studies on user acceptance (34–37).

We define the observed variables of user experience as the service quality of the Robo-taxi service by stage. We use the service quality evaluation values for a total of four stages: call, pick-up, traveling, and drop-off. The call stage is a process in which the participant enters a destination using a mobile phone app and finds a taxi. In the pick-up stage, when a taxi is dispatched through the app, the taxi's location is displayed on the app. The participant looks at the map, finds the location, recognizes the taxi, and boards. The traveling stage is a process in which the participant travels from the origin to the destination in a robot taxi. The drop-off stage is the process by which the taxi arrives near the destination, pulls over to the side of the road, and the occupants get off safely. The detailed evaluation method is described in section 4.3. We also selected overall satisfaction, intention to use, and willingness to pay (WTP) as the observed variables of user acceptance.

First, overall satisfaction refers to overall satisfaction after experiencing the Robo-taxi service. Satisfaction has been used as an important predictor variable of user behavioral intention (38), and studies related to public transportation services have mainly analyzed factors affecting satisfaction to increase user acceptance (e.g., taxi service (29), subway service (39), and railway service (40)).

Second, intention to use evaluates whether there is a plan to use Robo-taxis in the future and represents the meaning of user acceptance. Choi and Ji (41) argued that usefulness and trust were crucial factors for the intention to use self-driving vehicles, explaining the user's adoption factors of self-driving vehicles. In addition, many studies have analyzed ways to increase the intention to use self-driving vehicles (9,25,41).

Lastly, WTP is a variable for the reasonable price of the Robo-taxi service. Price perception and price acceptance play a vital role in affecting the user's consumption and post-consumption process (42–44). Price is also a determinant of value perception (45).

Based on these, we establish the following hypothesis:

**H1a:** The user experience of the Robo-taxi service will affect user acceptance.

**3.2. Effect of user emotion on user acceptance**

Studies on the acceptance of self-driving vehicles and user emotion have mainly covered specific emotions related to safety, such as anxiety and trust. Hohenberger et al. (9) showed that anxiety has a negative effect on the willingness to use self-driving vehicles. Choi and Ji (40) demonstrated that trust was a decisive factor in the intention to use self-driving vehicles. Similarly, Kaur and Rampersad (41) found that trust and performance expectations were decisive factors in the adoption of self-driving vehicles. Stanton and Young (46) considered the psychological variables related to driving automation, such as feedback, locus of control, mental workload, driver stress, situational awareness, and mental representation.

Extensive research on the relationship between user emotion and user acceptance has been conducted in other service industries (47). According to Ali et al. (48), customer satisfaction and price acceptance were affected when users felt lots of positive emotions through service experience. Lee et el. (49) investigated the effect of customers' positive and negative emotions on satisfaction and brand loyalty. Grace and O'Cass (50) revealed that service experience, emotion, satisfaction, and brand attitude were related.

Most studies find positive emotions, such as trust, built up user acceptance, whereas negative emotions such as anxiety and stress negatively affect user acceptance. We investigate which emotions greatly affect user acceptance among the detailed positive and negative emotions that the user feels through the experience of the Robo-taxi service. The 24 detailed emotions used as the observed variables are presented in Section 4.3 and based on these, we establish the following hypotheses.

**H2a:** Positive emotions regarding Robo-taxi technology will have a positive effect on user acceptance.

**H2b:** Negative emotions regarding Robo-taxi technology will have a negative effect on user acceptance.

# 4. Data

This section presents the vehicle and service configuration used in the experiment (Section 4.1), the experimental path (Section 4.2), the survey design (Section 4.3), and the participants (Section 4.4). We used data obtained from the Robo-taxi field test studies (21–22).

**4.1. Robo-taxi vehicle and service configuration**

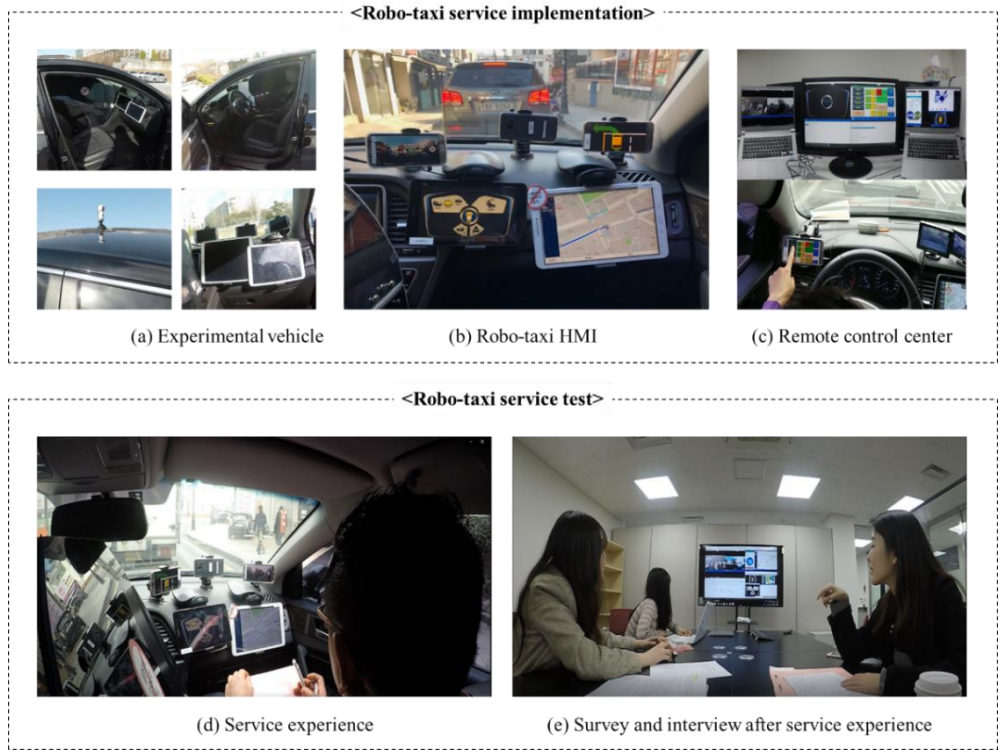

Figure 1. (a) From the upper left, front passenger seat, driver's seat, and the 360-degree camera attached in front of the passenger seat, the view seen when sitting in the passenger seat. (b) View from the passenger seat. There are three mobile phones used for 360-degree camera images, for recording the participant's images, and for directions. Of the two tablets, the left one is an app for communication with and the operation of the Robo-taxi, and the right one is for navigation. (c) (top) Robo-taxi control center environment, (bottom). Photo of controlling voice guidance through the app from the driver's seat. (d) The participant experiencing Robo-taxi in the field. (e) Interview with the participant while watching an experiment video after the Robo-taxi service (22).

It is difficult to provide the Robo-taxi service in a complex city center environment with current self-driving technology, and there are no laws and regulations in Korea for self-driving vehicles to drive on general roads. Therefore, we implemented the Robo-taxi service using the Wizard of OZ method, which is a way of getting participants to believe that they are using real services, but in reality, experimenters manually play the role of the automation system (51–52), completely blocking the driver's seat of the vehicle with partitions. The driver boarded and drove during the experiment, and by installing cameras, microphones, and speakers, the situation of the vehicle and participants was controlled in real-time from a remote control tower. We explained to the participant that a safety guard is in the driving seat, and she will control the Robo-Taxi manually only in case of emergencies. During the trip, participants were able to change the driving mode through the tablet's interface and voice recognition function (e.g. commands such as 'turn on the radio' or 'open the window'). When the remote control tower recognized the participant's command, the remote control tower delivered commands to the driver via a telephone connection, and the driver directly controlled the car to conduct the experiment as if the Robo-taxi worked. The voice communicated with the participant by machine voice using Text-to-Speech (TTS), and when driving, the handling was smoothed and sudden acceleration and sudden stops were avoided, making it feel more like a Robo-taxi. The survey confirmed that all participants believed that they were fully autonomous driving after experiencing the service. Figure 1 shows the Robo-taxi service implementation and the test scenes. The details of the experiment are described by Yoo et al. (22).

### 4.2. Robo-taxi service path

The experiments were conducted in the downtown areas of Daejeon and Seoul, Korea. The travel distance was approximately 7 km, and the travel time was approximately 30 min. The details and path of the Daejeon experiment are described in Kim et al. (21) and the Seoul experiment can be found in Yoo et al. (22). The experimental path was a mixture of a quiet and wide road, many vehicles and floating populations, and narrow alleys and steep hills for vehicles to pass. This path gave the participants experiences on not only comfortable situations but also uncomfortable ones. Beyond the unsafe scenarios such as sudden pedestrians that actually occurred, we artificially set up unexpected scenarios during the experiment, such as putting obstacles on the road or sounding the accident occurrence alarm which are likely to happen during Robo-taxi service.

### 4.3. Survey and interview by service stage

The participants received pre-guidance during the experiment. The participants returned to the laboratory for the post-survey and interview after experiencing the four stages of the Robo-taxi service (i.e., call, pick-up, traveling, drop-off). The survey questions are presented in Appendix A.

First, the participants performed a quantitative test on a 7-point Likert scale with the evaluation factors shown in Table 1 for each service stage. We selected appropriate evaluation factors for the Robo-taxi service by referring to those used in evaluating transportation systems in previous studies (39–40,53–55). The 1st level is to evaluate the service quality for each stage, and the evaluation factors corresponding to the 2nd level are detailed evaluation items that are likely to affect service quality. Table 2 presents the definitions of evaluation factors.

In addition, the overall satisfaction and intention to use the service were surveyed on a 7-point Likert scale. For the WTP question, there are seven options regarding how much participants were willing to pay relative to a manned taxi: less than 50%, 50 %–74%, 75–99%, 100%, 101 %–124%, 125–149%, and 150% or more.

Table 1. Perceptual evaluation factors by stage

| Stage | Evaluation factor | |
|---|---|---|
| | 1st level | 2nd level |
| Call | Service quality | Reliability, Promptness, Predictability, Information, Kindness, Convenience |
| Pick-up | Service quality | Reliability, Safety, Predictability, Information, Accessibility, Punctuality, Kindness, Communication, Confirmation |
| Traveling | Service quality | Reliability, Speed, Ride comfort, Safety, Predictability, Information, Kindness, Communication, Pleasantness, Convenience, Comfort |
| Drop-off | Service quality | Reliability, Safety, Predictability, Information, Accessibility, Punctuality, Communication, Kindness |

Table 2. Definition of quantitative evaluation factors (21)

| Evaluation factor | Description |
|---|---|
| Service quality | Felt satisfaction at each stage as integrated evaluation covering all other evaluation factors below. |
| Reliability | Felt that the service was reliable. |
| Predictability | It was possible to predict what had to be done. |
| Information | Necessary information was received properly. |
| Kindness | Felt kindness in the service. |
| Safety | Felt it was safe. |
| Communication | Communication with the taxi was satisfactory. |
| Accessibility | The taxi came to the desired place. |
| Punctuality | The taxi arrived at the predicted time. |
| Convenience | It was convenient and easy to use. |
| Promptness | The service was carried out promptly. |
| Confirmation | It was easy to identify my taxi. |
| Speed | The speed was appropriate. |
| Ride comfort | The ride was smooth and comfortable. |
| Pleasantness | Felt pleasant in the taxi. |
| Comfort | Felt comfortable psychologically. |

In addition to the evaluation by service stage, after experiencing the entire service, we conducted a quantitative survey that used a 7-point Likert scale for detailed emotions felt by the participants throughout the Robo-taxi service. Emotions, which were derived through brainstorming and a semantic differential method, were divided into positive and negative emotions. Table 3 shows the 12 positive and 12 negative emotions used in the evaluation.

Table 3. Emotion evaluation index (21)

| Type | Emotions |
|---|---|
| Positive | Convenient, comfortable, familiar, safe, reliable, excellent, simple, sophisticated, ingenious, trendy, efficient, new |
| Negative | Nervous, uncomfortable, afraid, unpleasant, annoying, disappointing, stuffy, tiresome, complicated, dull, strange, frustrating |

**4.4. Participants**

We recruited a total of 71 participants online for the field test; 43 and 28 people participated in the experiment in Daejeon and Seoul, respectively. By gender, 45% were men and 55% were women. By age, 8% were in their teens, 65% in their twenties, 14% in their thirties, 7% in their forties, and 6% in their fifties. When we examined the frequency of using taxis for the participants, 42% used taxis less than once a week, 28% used once a week, 24% used 2–3 times a week, 4% used 4–6 times a week, and 2% used every day. The participants were paid 20,000 Korean won per hour to participate in the experiment. The entire experiment and interview took approximately three hours for each participant. We checked the reliability of the responses by asking what percentage of Robo-taxi driving was completed during driving without the intervention of safety personnel. As a result, 74% responded that the Robo-taxi seemed to have been running fully autonomously without the intervention of safety personnel, and the rest replied that it seemed to have been running partially autonomously with some intervention by safety personnel.

# 5. Models and Results

We performed path analysis and SEM using SPSS Amos 23 (56) based on experience data from 71 users collected through the field test. We analyzed the significance of the hypotheses defined in Section 3. Each sub-section describes each of the three models. Model A finds the important factors of each stage that affect overall satisfaction by using path analysis (Section 5.1). Model B is a model that examines the relationship between user experience and user acceptance using SEM (Section 5.2). Model C analyzes potential emotions that significantly influence user acceptance using SEM (Section 5.3).

**5.1. Model A: path analysis for overall satisfaction and service quality in each stage**

We performed path analysis to identify crucial evaluation factors by boarding stage that significantly affected overall satisfaction, which was a crucial factor for user acceptance.

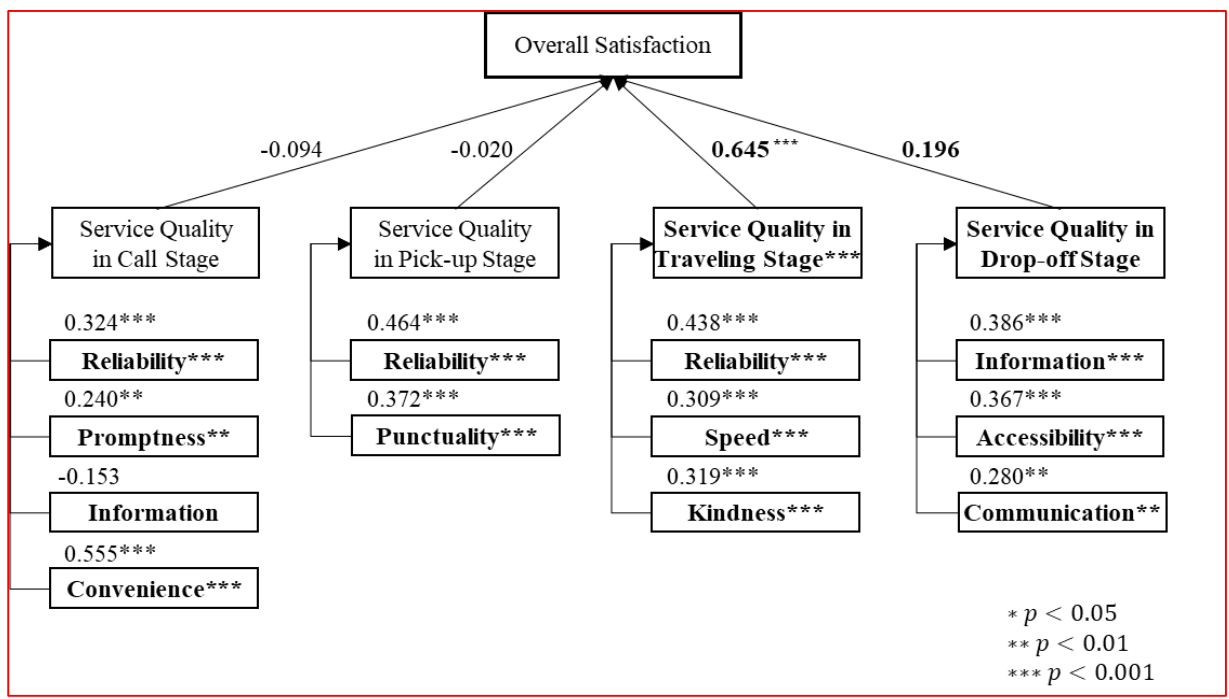

Figure 2. Path analysis model of overall satisfaction and service quality by service stage (Model A)

Figure 2 shows the results of path analysis. Path analysis is a multiple regression statistical analysis that examines the relationships between independent variables. By specifying the relationship between independent variables with arrows, we can analyze whether the independent variable affects the dependent variable (57). the evaluation factors in bold indicate significant or less significant, but still influential factors. When looking at multicollinearity between sub-variables and service qualities of each stage, all of the $R^2$ values were less than 0.9, and variance inflation factors (VIF) were less than 10; thus, no multicollinearity was found. The results showed that service quality in the traveling stage was significant ($p < 0.001$), and service quality in the drop-off stage was significant ($p < 0.1$). The estimates of service quality in the traveling and the drop-off stages were 0.643 and 0.196, respectively. That is, the effect of service quality on the traveling stage was the highest. Table 4 presents the path analysis results, which were significant at $p < 0.1$. The reason for the relaxation of the threshold is that the p-value was slightly high, but it was judged to be a sufficiently meaningful factor when looking at the interview results together.

Regarding service quality in the traveling stage, reliability and speed were significant, and the effect was in the order of reliability and speed. Regarding service quality in the drop-off stage, accessibility and information were significant. Of these, accessibility had the highest effect. Kindness in the traveling stage and communication in the drop-off stage did not satisfy p<0.05, with a narrow margin, but they were judged as important factors due to a large estimate.

The calculation results of the effect of the crucial factors on overall satisfaction with service quality in the traveling stage showed that reliability was 0.283 (=0.438x0.645), kindness was 0.206 (=0.319x0.645), and speed was 0.199 (=0.309x0.645). In the drop-off stage, information was 0.076 (=0.386x0.196), accessibility was 0.072 (=0.367x0.196), and communication was 0.055 (=0.280x0.196). Therefore, the factors that have the largest effect on overall satisfaction are reliability, speed, kindness, accessibility, information, and communication. The fact that service quality in the traveling stage was significant was also true in Model B in Section 5.2. However, path analysis revealed that reliability, speed, and kindness, which are the 2$^{nd}$ level evaluation factors, are crucial.

### 5.2. Model B: SEM for user experience and user acceptance

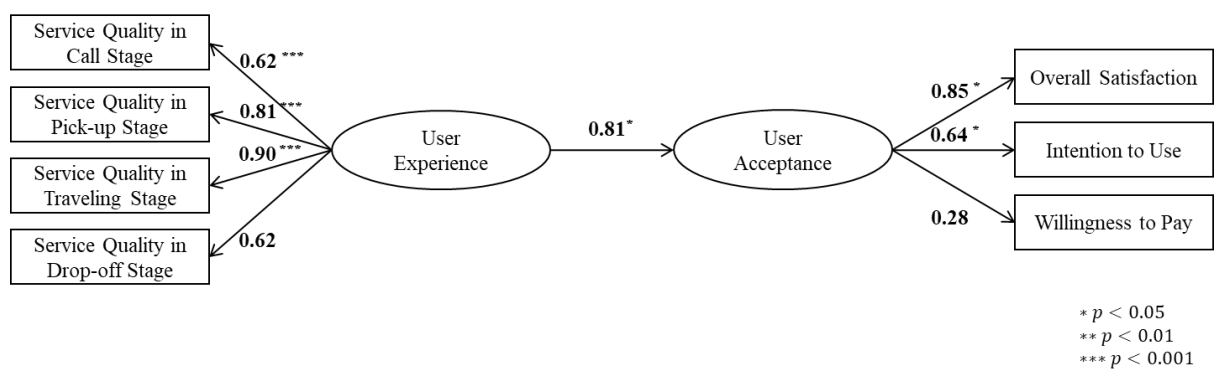

Figure 3. SEM for the relationship between user experience, user acceptance, and demographic (Model B)

Model B in Figure 3 assumes that user experience is related to user acceptance. The service quality of each stage of the Robo-taxi service explains the user experience. The data collected through the quantitative survey conducted after the experiment, that is, overall satisfaction regarding the service, intention to use Robo-taxi, and WTP, explained the potential variable, user acceptance. In the initial model, demographic factors such as gender, age, and usage frequency were added to the model, but all of these observed variables were insignificant and thus were excluded. The initial demographic model is presented in Appendix B1.

Model B can be represented as the following equations:

$$\mathbf{X} = \boldsymbol{\lambda}_x \xi + \boldsymbol{\delta}$$

$$\mathbf{Y} = \boldsymbol{\lambda}_y \eta + \boldsymbol{\varepsilon} \qquad (1)$$

$$\eta = \Gamma \xi + \zeta$$

where $\mathbf{X}$ is the observed exogenous variables indicating service quality. $X_1$ is service quality in call stage, $X_2$ is service quality in pick-up stage, $X_3$ is service quality in traveling stage, and $X_4$ is service quality in drop-off stage. $\xi$ is the latent exogenous variable indicating user experience. $\boldsymbol{\lambda}_x$ is the path from user experience $\xi$ to service quality $\mathbf{X}$. $\boldsymbol{\delta}$ is the errors of service quality $\mathbf{X}$. $\mathbf{Y}$ is the observed endogenous variables. $Y_1$ is overall satisfaction, $Y_2$ is intention to user, and $Y_3$ is willingness to pay. $\eta$ is the latent endogenous variable indicating user acceptance. $\boldsymbol{\lambda}_y$ is the path from user acceptance $\eta$ to $\mathbf{Y}$. $\boldsymbol{\varepsilon}$ is the errors of $\mathbf{Y}$. $\Gamma$ is the path from user experience $\xi$ to user acceptance $\eta$. $\zeta$ is the error of user acceptance $\eta$.

The results of the goodness of fit test of the model showed that the normed-fit index (NFI) = 0.84, the goodness of fit (GFI) = 0.85, root mean square residual (RMR) = 0.09, and root mean square error of approximation (RMSEA) = 0.13. RMSEA deviated slightly from the threshold criterion, yet showed an acceptable fit. NFI is judged to be appropriate at 0.8-0.9 or higher (58), and GFI is judged to be appropriate at 0.8-0.9 (59). RMR is judged to be appropriate at 0.08, or lower (60), and RMSEA is judged to be appropriate at lower than 0.08 (61). In Figure 4, the numbers above the arrows indicate the standardized regression weights for each relationship.

User experience was significant ($p < 0.05$) for user acceptance and had a positive correlation with a magnitude of 0.81. Service quality in the call, pick-up, traveling, and drop-off stages were all $p < 0.001$, which significantly explained the potential variable, user experience. Intention to use was $p < 0.05$, and overall satisfaction was $p < 0.05$, which significantly explained the potential variable. Of the observed variables of user experience, service quality in the traveling stage was the highest (0.90) when examining the estimate in standardized regression weights. For the rest, the regression coefficient was high in the order of service quality in the pick-up stage with 0.81, in the call stage with 0.62, and in the drop-off stage with 0.62. This indicates that the most important factor in the user experience is the service quality in the traveling stage. Of the observed variables of user acceptance, intention to use was 0.64, WTP was 0.28, and overall satisfaction was 0.85.

In conclusion, in the case of hypothesis H1a, user experience was positively correlated with user acceptance; thus, the hypothesis was accepted.

**5.3. Model C: SEM for positive and negative emotions and user acceptance**

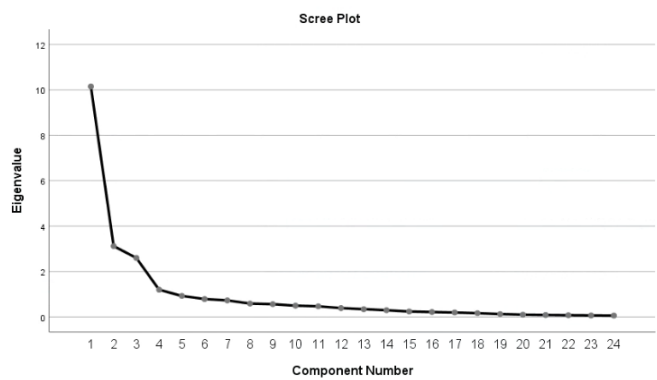

Figure 4. Scree plot of 24 emotional factors

We evaluated a total of 24 emotions and used exploratory factor analysis (EFA) to select typical emotions. To determine the number of factors, we drew a scree plot of the 24 emotion factors. As shown in Figure 4, the component numbers between two and four indicate the elbow. We tested the SEM by reducing the factors from four to two in sequence. When the factor was reduced to 4 in the SEM, the p-value of one latent variable was not significant. When the factor was reduced to 2 in the SEM, the total cumulative explanatory amount of PCA was insufficient. (Results of factors 2 and 4 PCA and the model are presented in Appendix B5-6.) For this reason, we fixed components to three and performed principal component analysis (PCA) using varimax factor rotation. The varimax factor is one of the methods used to rotate a factor in factor analysis. The varimax factor stands for "Variance is maximized" which maximizes the variance of factors so that factors can be interpreted (62). The results are shown in Table 5.

Table 5. Results of PCA of 24 emotional factors

| Factor | Component | | |
|---|---|---|---|
| | 1 | 2 | 3 |
| New | **0.885** | | |
| Ingenious | **0.876** | | |
| Trendy | **0.845** | | |
| Excellent | 0.798 | | |
| Sophisticated | 0.711 | | |
| Efficient | 0.697 | | |
| Simple | 0.673 | | |
| Familiar | 0.632 | | |
| Convenient | 0.594 | | |
| Reliable | 0.582 | | |
| Unpleasant | | **0.815** | |
| Disappointing | | **0.790** | |
| Annoying | | **0.789** | |
| Strange | | 0.772 | |
| Tiresome | | 0.756 | |
| Complicated | | 0.663 | |
| Dull | | 0.619 | |
| Frustrating | | 0.485 | |
| Stuffy | | 0.414 | |
| Uncomfortable | | | **0.886** |
| Nervous | | | **0.844** |
| Afraid | | | **0.843** |
| Safe | | | -0.767 |
| Comfortable | | | -0.579 |

The Kaiser–Meyer–Olkin measure (KMO) represents the degree to which correlations between variables are explained by other variables. KMO >0.7 is judged to be good, our result is good at 0.848. Bartlett's test of sphericity determines whether the use of factor analysis is appropriate if the p-value is less than 0.05. In this case, the p-value was appropriately set to 0.000. The total cumulative explanatory amount of PCA was 66.11%, indicating that the explanatory power was sufficient for the factors to be divided into three components. We selected the top three factors from the results of the factor analysis. New, ingenious, and trendy were selected as the final factors for positive emotions, while unpleasant, disappointing, annoying, uncomfortable, afraid, and nervous were selected as negative emotions. Positive emotions were represented as a potential variable called "cutting-edge". Unpleasant, disappointing, and annoying emotions were represented as a potential variable called "bothersome". Uncomfortable, nervous, and afraid emotions were represented as a potential variable called "apprehensive". We also tested various ways to add demographic data to the model, but the results were not significant.

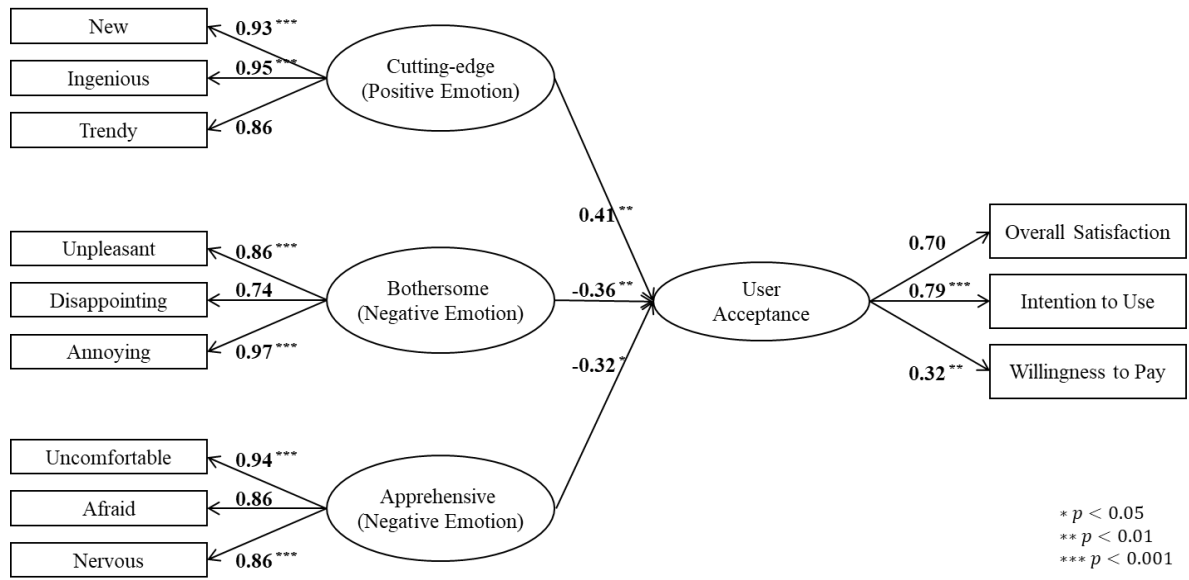

Figure 5. SEM showing the relationship between positive and negative emotions and user acceptance (Model C)

Model C can be represented as the following equations:

$$\mathbf{X} = \boldsymbol{\lambda}_x \boldsymbol{\xi} + \boldsymbol{\delta}$$

$$\mathbf{Y} = \boldsymbol{\lambda}_y \eta + \boldsymbol{\varepsilon} \qquad (2)$$

$$\eta = \boldsymbol{\Gamma}\boldsymbol{\xi} + \zeta$$

where $\mathbf{X}$ is the observed exogenous variables indicating emotions. $X_1$ indicates new, $X_2$ ingenious, $X_3$ trendy, $X_4$ unpleasant, $X_5$ disappointing, $X_6$ annoying, $X_7$ uncomfortable, $X_8$ afraid, and $X_9$ nervous emotions. $\boldsymbol{\xi}$ is the latent exogenous variables. $\xi_1$ represents cutting-edge, $\xi_2$ bothersome, and $\xi_3$ apprehensive emotions. $\boldsymbol{\lambda}_x$ is the path from $\boldsymbol{\xi}$ to emotion $\mathbf{X}$. $\boldsymbol{\delta}$ is the errors of $\mathbf{X}$. $\mathbf{Y}$ is the observed endogenous variables. $Y_1$ is overall satisfaction, $Y_2$ is intention to user, and $Y_3$ is willingness to pay. $\eta$ is the latent endogenous variable indicating user acceptance. $\boldsymbol{\lambda}_y$ is the path from user acceptance $\eta$ to $\mathbf{Y}$. $\boldsymbol{\varepsilon}$ is the errors of $\mathbf{Y}$. $\boldsymbol{\Gamma}$ is the path from $\boldsymbol{\xi}$ to user acceptance $\eta$. $\zeta$ is the error of user acceptance $\eta$.

The model in Figure 5 is based on the assumption that the emotional factors representing cutting-edge, bothersome and apprehensive are related to user acceptance. The results of the goodness of fit test showed that NFI = 0.91, GFI = 0.90, RMR = 0.10, and RMSEA = 0.04, thus showing acceptable fitness. The RMR deviated slightly from the threshold criterion, yet showed an acceptable fit. The threshold is described in section 5.2. In Figure 6, the numbers above the arrows indicate the standardized regression weight of each relationship, which was rounded to the third decimal place.

The cutting edge was significant ($p < 0.01$) for user acceptance and had a positive correlation with the estimate of 0.41. The bothersome was significant ($p < 0.01$) for user acceptance and had a negative correlation with the estimate of -0.36. Apprehensive was significant ($p < 0.05$) for user acceptance and has a negative correlation with the estimate of -0.32. Of the observed variables of cutting-edge, both new and ingenious were significant with $p < 0.001$, and both factors have a similar regression coefficient of 0.93 and 0.95, respectively. Of the observed variables of bothersome, both unpleasant and annoying were significant with $p < 0.001$, and had a regression coefficient of 0.86 and 0.97, respectively. Of the observed variables of apprehensive, both uncomfortable and nervous were significant with p <0.001, and had a regression coefficient of 0.94 and 0.86, respectively. Of the observed variables of user acceptance, intention to use, and WTP were significant (p <0.001 and p <0.01, respectively). Intention to use had a significant effect of 0.79, and overall satisfaction was 0.70, yet the effect of WTP was relatively small at 0.32.

In conclusion, in the case of hypothesis H2a, cutting-edge, which is a typical positive emotion for the Robo-taxi, was positively correlated with user acceptance; thus, the hypothesis was accepted. In the case of hypothesis H2b, bothersome and apprehensive, which is a typical negative emotion towards the Robo-taxi, was negatively correlated with user acceptance; thus, the hypothesis was also accepted.

# 6. Discussion

We discuss the quantitative results derived from Section 5 and analyze the qualitative results of the field test interviews to derive meaningful insights.

### 6.1. User experience and user acceptance

First, the results of model B showed that positive experience with the Robo-taxi service affected the improvement of user acceptance. This indicates that there is a need for a strategy that enables as many customers as possible to experience the service in advance at the beginning of the Robo-taxi service launch. Since user acceptance is high for those who have positive user experiences at the beginning of the service, they are highly likely to continue using the service. Therefore, it seems that various early marketing strategies, such as free trials, are critical. The additional field test results also support these results. Positive responses to questions about using Robo-taxis increased after the field test compared to before the field test. The interview results regarding the intention to use Robo-taxis before and after the field test are as follows. Before the field test, 39% of the participants responded that they would use Robo-taxi as soon as it was commercialized, and 61% responded that they would use it when it was considered safe enough after its commercialization. After the field test, 96% of the participants responded that they would use Robo-taxi as soon as it was commercialized, while 4% responded that they had no intention to use it or only use it when it became safe after its commercialization. Interestingly, these 4% had experienced more than one traffic accident and when asked how many years it would be safe for Robo-taxis, they all answered "after 2050." A previous study, Rödel et al. (24), also found that prior experience of self-driving vehicles had a positive effect on user acceptance and experience.

Second, in both Models A and B, the effect of the traveling stage experience was the highest. The interview results also showed that the travel stage was the most crucial. After the field test, of the answers to the question "What was good about the Robo-taxi compared to general taxis?" The differentiation of the traveling stage between them was good and accounted for 97%. In particular, approximately 50% were about the convenience provided by unmanned services during travel. They were satisfied with the service in that they did not have to talk to the driver, the environment was private, and there was no body or cigarette odor inside the vehicle. Some responses are quoted below.

*"I loved being able to do whatever I wished with nobody around me." (p05)*

*"I didn't have to talk to the taxi driver and there was no burden." (p63)*

*"I felt comfortable without emotional discomfort in the absence of the driver." (p68)*

Approximately 30% of the reasons why the Robo-taxi was good were that AI was driving, driving was safe, and riding quality was good.

*"I felt safe because it followed the traffic rules. It was good to be quiet'." (p08)*

*"It was good to keep the appropriate distance from the car in front. I thought it was safe, and it seemed to follow the rules well." (p53)*

There were other views about freedom of action, such as the possibility of consuming food, reclining and sleeping, and answering calls. This indicates that focusing on the development and improvement of the service for the traveling stage can greatly contribute to the improvement of technology acceptance.

Third, in Model B, of the variables representing user acceptance, WTP showed relatively low significance. Additionally, the Pearson correlation coefficient between the demographic variables (i.e., age, gender, and usage frequency) and WTP was not significant. Of the participants who responded positively to the intention to use, some indicated that they were willing to pay more because Robo-taxi was a new technology, whereas others felt that the fare should be lowered because there was no driver. Therefore, there are different views regarding WTP. Several previous studies argued that it was possible to reduce the transportation cost because there was no need for a driver; accordingly, the user expected the price advantage of the Robo-taxi (44–45,63). Some responses are quoted below.

*"The fare initially seems to be similar to manned taxis. However, if a person drives directly, labor charges apply because it is a service job; thus, Robo-taxi seems to have a different pricing policy because it simply uses machines." (p44)*

*"I will use it more frequently when the fare is lowered and commercialized." (p45)*

*"Just like expressway buses offering luxury or premium amenities according to the class, if the Robo-taxi had such features, I would use it." (p69)*

### 6.2. Cutting-edge and user acceptance

In Model C, the cutting-edge, which represents positive emotions, affects user acceptance. The cutting edge can be explained by emotions such as new, ingenious, and trendy. With an increase in the newness (freshness) users feel when they experience Robo-taxis, user acceptance can increase. Therefore, the differentiation between Robo-taxi and traditional taxi services should be maximized.

The interview results revealed that the participants thought highly of the cutting edge of the new technology itself. Although they felt that Robo-taxis were not perfect, they expected Robo-taxis to make them more comfortable in the future. Some responses are quoted below.

*"I felt it was more convenient. It was good to automatically check if the seat belt was fastened." (p11)*

*"I thought it was convenient and I would do more activities if unmanned taxis were commercialized." (p50)*

*"Above all, I liked it because I didn't have to tell the driver about my destination, there was no refusal of passengers, and it was good to display the navigation paths on the big screen." (p69)*

### 6.3. Service apprehension and user acceptance

In Model C, of the negative emotions, we confirmed that apprehension regarding new technology affected user acceptance. We defined apprehensive as emotions such as uncomfortable, afraid, and nervous. To increase user acceptance, service apprehension needs to be addressed first. The interview results revealed that there was some apprehension towards using Robo-taxi since it was a new technology that the participants had never experienced before. In the early traveling stage, they felt uncomfortable about whether it worked well or was safe without a driver, but generally, in the later traveling stage, they responded that they felt relatively less uncomfortable, afraid, and nervous. Some responses are quoted below.

*"I had doubt on how reliable this technology could be. I wondered if Robo-taxis could deal with unexpected situations." (p47)*

*"I was terrified. I felt more uncomfortable because I was scared at the beginning." (p65)*

*"I felt uncomfortable at the beginning, but I was okay from the middle of the service. If passengers are provided with a preliminary explanation/coping method for some specific situations before boarding an unmanned taxi, it will likely reduce discomfort." (p67)*

### 6.4. Bothersome and user acceptance

Regarding Model C, bothersome, negative emotions as an observation variable affected user acceptance. We defined bothersome as unpleasant, disappointing, and annoying. To increase user acceptance, Robo-taxis need to increase their 'human-free advantage', which makes the Robo-taxi better than regular taxis. Participants mentioned many advantages of not having a regular taxi driver. Participants did not prefer unfriendly taxi drivers, unwanted small talks with taxi drivers, taxi driver's body odor, or the smell of cigarettes. Thus, participants felt less negative emotions for bothersome for Robo-taxis than for regular taxis. Some responses are quoted below.

*"There was no need to talk to the taxi driver and there was no burden..."(p63)*

*"There was no burden or discomfort about the driver's attitude. For example, the smell of cigarettes from the body, unkind words, and actions" (p49)*

*"It was comfortable because I didn't have to care about the driver." (p70)*

### 6.5. Service apprehension and level of autonomy

After the field test, the participant was asked the following question: "How much do you think the driverless robot taxi drove without the help of the safety guard while driving? Please tell me as a percentage. Why do you think so?" We found that there was an inverse correlation between participants' apprehension and the level of autonomy. The apprehensiveness decreased when participants believed that the Robo-taxi was operated with full automation, and the apprehensive tended to increase when the participant had doubts about the autonomous driving technology. The correlation results are presented in Table 6.

Table 6. Pearson correlation coefficient of apprehension with the level of autonomy

|  | Pearson correlation | p-value |
|---|---|---|
| Nervous | -0.322 | 0.095 |
| Uncomfortable | -0.380 | 0.046 * |
| Afraid | -0.366 | 0.056 |

*$p<0.05$

Below is an interview in which the participant answered why they believed that the Robo-taxi drove themselves.

*"The accelerator pedal, brake, steering wheel, etc. did not sound, so I thought it was autonomous driving. It was nice to go slowly and keep the signals well." (p47)*

*"At first I thought the driver was driving, but when I saw the rearview mirror covered, I thought the driver wasn't driving." (p68)*

### 6.6. Reliability, speed, and kindness while traveling

Model B showed that traveling-stage experience had a significant effect on technology acceptance. Reliability, speed, and kindness had a significant effect on the traveling-stage experience in Model A. In previous research, reliability was regarded as the most crucial factor when experiencing Robo-taxi (11); in addition, as with the studies that demonstrated that the majority of experiment participants preferred the driving style of Robo-taxi to be similar to their driving style (15), it is expected that user acceptance can be increased if the optimization of the user-customized speed increases satisfaction regarding the speed factor. In particular, reliability and speed were found to be related. Many users seemed to be very uncomfortable because they did not trust the new technology, and some users evaluated satisfaction according to speed. They felt differently even at the same speed, and some felt uncomfortable at high speeds, and vice versa. Some responses are quoted below.

*"I felt a little discomfort as this was my first time experiencing an unmanned taxi. I felt that the speed of the taxi was not completely consistent, and I noticed a bit of a sudden jerk when it changed lanes." (p13)*

*"The speed was reasonable, and it was comfortable and good like a normal taxi." (p55)*

*"The speed was a little slow compared to a normal car, and I felt a little discomfort about self-driving cars." (p7)*

*"It drove slowly in crowded alleys, so I felt safe." (p62)*

Regarding the interview results related to kindness, some participants felt that the voice from the systemized machine was kind in the absence of a driver, whereas there were opposing views that since it was an efficient system that aimed to get to the destination from a departure point, they did not feel any kindness. Since kindness turned out to be crucial in the traveling stage, which was most important in the Robo-taxi service, the way that a Robo-taxi could engage the user emotionally should also be considered. Some responses are quoted below.

*"I didn't have to worry about the driver's mood, a chat with the driver, and uncomfortable things." (p61)*

*"I didn't have to unwillingly talk to the driver, and it was good that all communication was done by machine." (p56)*

*"I believe that kindness is such a thing that if there was a driver, the driver could have a humorous chat with me or could greet me, but I was not able to share the emotion of kindness with an unmanned taxi. I just got in the car, but I am not sure if there is any element of kindness." (p69)*

### 6.7. Accessibility, information, and communication while dropping off

In Model A, in the drop-off stage, particular attention should be paid to accessibility, information, and communication to increase overall satisfaction. In the case of a manned taxi, the driver drops the passenger off flexibly by identifying traffic situations at their discretion. However, Robo-taxis drops passengers off only at the designated safe stop. Therefore, there is a need to develop an HMI for flexible dropping off in terms of accessibility. Passengers have an approximate destination, but they should be able to specifically ask Robo-taxis exactly where to get out around the destination. In the same context, Robo-taxis should be able to offer passengers a safe place to be dropped off so that they can choose where to be dropped. The interview results also revealed that it was inconvenient in terms of the accessibility of where to get out and not being able to specify this through information and direct communication. Some responses are quoted below.

*"I could not get out at the right place when I arrived. Voice recognition made it difficult for me to specifically control it." (p26)*

*"I felt uncomfortable when it stopped because of a car in front." (p12)*

*"It went smoothly. Before getting out, I asked about AI voice recognition for the expected time to be dropped off. I was satisfied with the information provided." (p68)*

*"I was a little confused when to get out." (p23)*

*"In general, when I ask a taxi driver to be dropped off at a desired place, the driver drops me off. However, it seemed impossible for an unmanned taxi to drop me off flexibly. From the first call, it is necessary to determine the arrival place specifically. It would be inconvenient if the drop-off place was set in advance." (p10)*

*"When I wanted to get off near my destination, I couldn't convey my message to Robo-taxi." (p37)*

### 6.8. Heterogeneity

To examine the heterogeneity of the Robo-taxi service, we used a t-test to analyze the relationship between demographic (e.g., gender and age) and user acceptance (i.e., overall satisfaction, intention to use, and WTP). Overall satisfaction was significantly different depending on gender (i.e., p-value=0.004), where males had higher overall satisfaction than females (i.e., male: 5.75, female: 4.98, on average). This gender difference was also observed in previous studies (10). However, there was no significant relationship between gender and intention to use and WTP. In addition, no significant relationship was found between age and user acceptance. A more detailed analysis of gender and age differences is presented in Appendix B2.

### 6.9. Distribution and correlation of the responses

Looking at the distribution of service quality in each stage, the average service quality of the pick-up stage was the highest. This is because the process was similar to that of using a regular taxi according to the interview results. A detailed analysis is presented in Appendix B3.

In addition, as a result of the correlation analysis of each stage factor, predictability was directly correlated with reliability at all stages. Among them, communication between Robo-taxis and people was important to finding a Robo-taxi in the pick-up stage, and sufficient information about Robo-taxis was important in the traveling stage. In the drop-off stage, communication was important for the Robo-taxi to safely drop off passengers near the destination. Information and communication are also correlated with kindness; consequently, Robo-taxis should pay attention to information and communication in each stage to make passengers feel kindness and reliability. The detailed analysis results are presented in Appendix B4.

## 7. Conclusion

Unlike previous user experience studies on self-driving vehicles that mainly depended on surveys and simulators, this study was based on a survey and in-depth interview data obtained after participants experienced the Robo-taxi service operating in the actual city center. Based on the collected quantitative data, we built structural equation models and a path analysis model and analyzed them by comparing them with the in-depth interview results, which were qualitative data. Furthermore, by analyzing the relationship between user experience and user acceptance, we suggest a way to further improve the Robo-taxi service.

First, for the results of Model A, we performed path analysis for the relationship between the 34 evaluation factors by service stage and overall satisfaction. Service quality in the traveling and drop-off stages was found to have a significant effect on overall satisfaction. Reliability, speed, and kindness were found to be crucial factors in the traveling stage, and accessibility, information, and communication were found to be crucial factors in the drop-off stage. Second, the results of Model B, which showed the relationship between user experience and user acceptance, proved that user experience had a significant effect on user acceptance. Service quality in the traveling stage had the largest effect on user experience, and overall satisfaction had the largest effect on user acceptance, whereas WTP had a relatively low effect. Third, Model C, which analyzed 24 emotional factors and the relationship between these factors and user acceptance, showed that cutting-edge was selected as the typical emotion that had a positive relationship with user acceptance, whereas bothersome and apprehensive were selected as typical emotions that had a negative relationship. New, ingenious, and trendy relate to cutting-edge and unpleasant, disappointing, and annoying relate to bothersome and uncomfortable, afraid, and nervous relate to apprehensive.

Based on the results, we also suggest a robot-taxi service design and marketing. Since the initial service experience turned out to be crucial, there is a need for a strategy to provide as many customers as possible with service experience in the early stage of the Robo-taxi launch. Moreover, since cutting-edge has a significant effect on user acceptance, the differentiation between Robo-taxi and conventional taxi services should be maximized. Apprehension, a negative emotion, also affects user acceptance; thus, the issue of low reliability should be addressed more than anything else. Since reliability, speed, and kindness were found to be crucial in the traveling stage, it is necessary to provide an optimized speed service for an individual user and for in-vehicle AI to approach the user emotionally. In the drop-off stage, because accessibility, information, and communication are crucial, there is a need to develop an HMI for flexible drop-off (e.g., virtual stop (21)).

This study makes the following contributions. First, it conducts an SEM analysis based on survey data on the Robo-taxi user experience. The findings of previous studies were mainly based on simple surveys or simulator-dependent data, and there was a lack of SEM study cases based on real user experience. Second, we simultaneously analyzed the quantitative SEM and qualitative interview results. By identifying crucial relationships between the factors based on the SEM results and deriving the basis for them from the interview results, we improved the reliability of the model analysis results and performed a balanced analysis. Third, through the analysis results, we suggest guidelines for the design and marketing of future Robo-taxi services.

The limitations of this study and future research directions are as follows. First, owing to the physical limitations of the Robo-taxi field test, it was difficult to secure a statistically sufficient amount of user experience data. In future research, we will continue to increase the number of participants with diverse backgrounds. Participants' social status and income level will be collected and analyzed in detail. Second, we will collect Robo-taxi service data by considering various environments. For example, we will take night service situations where the demand for Robo-taxis is high, and service environments for the socially disadvantaged, such as the disabled and the elderly, into account. Third, only the correlation between user experience and user acceptance was examined in this study. To confirm causality, further research is needed to sufficiently increase the number of experimental data. In addition, we plan to capture the effect of unobserved heterogeneity due to unobserved factors (e.g., income and travel mode choice). Fourth, we plan to test unsafe driving scenarios separately, and analyze how these negative experiences affect user acceptance.

## Acknowledgments

This work was supported by National Research Foundation of Korea (NRF) grants funded by the Korean government [grant numbers 2017R1C1B2005266 and 2018R1A5A7025409].

**Declarations of interest: none**

# Appendix A: Survey questionnaire

**1. Pre-Survey & Interview**

*1.1. Survey*
1) Name: ______________
2) Gender
① Male ② Female
3) Age: ____
4) Occupation: ______________
5) Do you have any driving experience?
① None ② Less than 3 years   ③ 3-6 years        ④ 6-10 years       ⑤More than 10 years
6) How often do you drive?
① Less than once a week  ② Once a week   ③ 2-3 times a week  ④ 4-6 times a week   ⑤ Every day
7) How often do you use taxis?
① Less than once a week  ② Once a week   ③ 2-3 times a week  ④ 4-6 times a week   ⑤ Every day
8) How anxious are you when you ride a vehicle driven by other people?
Not at all         1 2 3 4 5 6 7      very anxious
9) How anxious are you when you ride a taxi?
Not at all         1 2 3 4 5 6 7      very anxious
10) When you take a taxi, where do you usually sit?
① Passenger seat ② Rear seats     ③ Different each time
11) Have you ever been injured in a traffic accident?
① None    ② Once     ③ Twice or more
12) Have you ever read or seen articles, books, or videos about autonomous vehicles?
① No   ② Yes
13) How safe do you think the current autonomous driving technology is compared to human driving?
<Not safe at all>  1  2  3  4  5  6  7  <very safe>
14) When do you expect autonomous driving technology will be safer than human driving?
    ① After 2020
    ② After 2030
    ③ After 2040
    ④ After 2050
    ⑤ After 2060

*1.2. Interview*
1) If you feel anxious when you ride a vehicle driven by other people, what is the reason?
2) If you feel anxious when you ride a taxi, what is the reason?
3) What would it look like if a Robo-taxi service becomes available? Please imagine.
4) What do you expect from Robo-taxi services?
5) Do you have any concerns over Robo-taxi services?
6) Would you like to use Robo-taxi services if they are available?

## 2. Post-Survey & Interview
*2.1. Survey*

1) Please evaluate each service step (1 point: not very much, 4 points: normal, and 7 points: very much)

a) Calling step

| Evaluation element | Description | Evaluation score | | | | | | |
|---|---|---|---|---|---|---|---|---|
| | | 1 | 2 | 3 | 4 | 5 | 6 | 7 |
| Reliability | The service was reliable. | | | | | | | |
| Predictability | I could predict what to do. | | | | | | | |
| Information | I received the necessary guidance information well. | | | | | | | |
| Kindness | The service was kind. | | | | | | | |
| Convenience | The service was convenient and easy to use. | | | | | | | |
| Promptness | The service was prompt. | | | | | | | |
| **Overall satisfaction** | **Overall, I was satisfied with the calling step.** | | | | | | | |

b) Boarding step

| Evaluation element | Description | Evaluation score | | | | | | |
|---|---|---|---|---|---|---|---|---|
| | | 1 | 2 | 3 | 4 | 5 | 6 | 7 |
| Reliability | The service was reliable. | | | | | | | |
| Predictability | I could predict what to do. | | | | | | | |
| Information | I received the necessary guidance information well. | | | | | | | |
| Kindness | The service was kind. | | | | | | | |
| Safety | I felt safe. | | | | | | | |
| Communication | Communication with the taxi was good. | | | | | | | |
| Accessibility | The vehicle came to the desired place. | | | | | | | |
| Punctuality | The vehicle arrived at the starting point at the expected time. | | | | | | | |
| Identification | It was easy to identify my taxi. | | | | | | | |
| **Overall satisfaction** | **Overall, I was satisfied with the boarding step.** | | | | | | | |

c) Traveling step

| Evaluation element | Description | Evaluation score | | | | | | |
|---|---|---|---|---|---|---|---|---|
| | | 1 | 2 | 3 | 4 | 5 | 6 | 7 |
| Reliability | The service was reliable. | | | | | | | |
| Predictability | I could predict what to do. | | | | | | | |
| Information | I received the necessary guidance information well. | | | | | | | |
| Kindness | The service was kind. | | | | | | | |
| Safety | I felt safe. | | | | | | | |
| Communication | Communication with the taxi was good. | | | | | | | |
| Convenience | The service was convenient and easy to use. | | | | | | | |
| Speed | The speed of the taxi was appropriate. | | | | | | | |
| Ride comfort | The ride comfort was good. | | | | | | | |
| Comfort | The taxi was comfortable. | | | | | | | |
| Relaxation | I felt relaxed. | | | | | | | |
| **Overall satisfaction** | **Overall, I was satisfied with the traveling step.** | | | | | | | |

d) Getting-off step

| Evaluation element | Description | Evaluation score | | | | | | |
|---|---|---|---|---|---|---|---|---|
| | | 1 | 2 | 3 | 4 | 5 | 6 | 7 |
| Reliability | The service was reliable. | | | | | | | |
| Predictability | I could predict what to do. | | | | | | | |
| Information | I received the necessary guidance information well. | | | | | | | |
| Kindness | The service was kind. | | | | | | | |
| Safety | I felt safe. | | | | | | | |
| Communication | Communication with the taxi was good. | | | | | | | |
| Accessibility | The vehicle arrived at the desired place. | | | | | | | |
| Punctuality | The vehicle arrived at the destination at the expected time. | | | | | | | |
| **Overall satisfaction** | **Overall, I was satisfied with the getting-off step.** | | | | | | | |

2) Please evaluate your feelings about the overall service.

Positive feelings (1 point: not very much, 4 points: normal, and 7 points: very much)

| Evaluation element | Evaluation score | | | | | | |
|---|---|---|---|---|---|---|---|
| | 1 | 2 | 3 | 4 | 5 | 6 | 7 |
| Convenient | | | | | | | |
| Comfortable | | | | | | | |
| Friendly | | | | | | | |
| Safe | | | | | | | |
| Reliable | | | | | | | |
| Outstanding | | | | | | | |
| Simple | | | | | | | |
| Refined | | | | | | | |
| Innovative | | | | | | | |
| Trendy | | | | | | | |
| Efficient | | | | | | | |
| Novel | | | | | | | |

[Negative feelings] (1 point: not very much, 4 points: normal, and 7 points: very much)

| Evaluation element | Evaluation score | | | | | | |
|---|---|---|---|---|---|---|---|
| | 1 | 2 | 3 | 4 | 5 | 6 | 7 |
| Nervous | | | | | | | |
| Anxious | | | | | | | |
| Afraid | | | | | | | |
| Unpleasant | | | | | | | |
| Irritated | | | | | | | |
| Disappointed | | | | | | | |
| Uncomfortable | | | | | | | |
| Troublesome | | | | | | | |
| Complex | | | | | | | |
| Bored | | | | | | | |
| Strange | | | | | | | |
| Embarrassed | | | | | | | |

3) Are you willing to pay for Robo-taxi services?

① Yes
② No

4) If you are willing to use Robo-taxis, what is your desired fare for Robo-taxis compared to conventional taxis?
① 50% or less
② 50-74%
③ 75-99%
④ 100% ( same as conventional taxis)
⑤ 101-124%
⑥ 125-149%
⑦ 150% or higher

*2.2. Introduction interview*

1) You have used a Robo-taxi, what was your overall feeling?
2) Compared to conventional taxis, what was good about the Robo-taxi?
3) Compared to conventional taxis, what was bad about the Robo-taxi?
4) When did you feel anxious and why?

*2.3. Video interview*

[Questions for each clicker situation were expected. ]

1) At this moment when you pressed the clicker, can you explain the situation?
2) At this moment, when you pressed the clicker, how can you rate your anxiety?
   (Not anxious at all     1 2 3 4 5 6 7     very anxious)
3) Why were you anxious?
    (e.g., mechanical functions of the Robo-taxi / service functions of the Robo-taxi / functions of the navigation system/interaction with the control tower)
4) When you were anxious, what responses or actions did you do? What actions do you usually do when you are anxious about a vehicle?
5) After pressing the clicker, what did you do to relieve anxiety?
6) Do you think that the same situation would be felt if you were in a conventional taxi? (How much anxiety is related to the robot taxi?)
    (My anxiety had nothing to do with the Robo-taxi / I would not have felt anxious if it had not been for the Robo-taxi / I have no idea ).
7) At that moment, what functions or services would have helped relieve anxiety?

*2.4. Finishing interview*

1) If you use a Robo-taxi next time, what activities will you do while driving?
2) What will you do if the Robo-taxi shows abnormal behavior, such as suddenly changing the driving path or error signals?
3) Please tell us if there were any inconveniences during the test.

**3. Contents Added to the 2nd Field Test**

*3.1. Survey*

Q1) How much were the following functions helpful in relieving anxiety? (1-7 point Likert scale)
    a) Navigation
    b) Blinkers
    c) 360° camera image
    d) Speed control function
    e) Horn function

      f) Emergency stop function
      g) Sleep mode
      h) AI voice mode

Q2) Now you have used the following functions: How much do you think they are necessary for Robo-taxis? (1-7 point Likert scale)
      a) Navigation
      b) Blinkers
      c) 360° camera image
      d) Speed control function
      e) Horn function
      f) Emergency stop function
      g) Sleep mode
      h) AI voice mode

Q3) Vehicle identification was performed during the boarding step in the following process. How helpful is this function in vehicle identification? (1-7 point Likert scale)
      Step 1: The horn is sounded twice in a small magnitude at a radius of 3 m.
      Step 2: Rear mirrors are unfolded at a 0.5-1 m radius.
      Step 3: The participant boards the robot taxi by pressing the app's door open button in the same way as before.

### 3.2. Interview

Q1) Please tell us why the above functions were helpful or not helpful in relieving anxiety and why they are necessary or not necessary for robot-taxis. Please indicate the items to be supplemented or added to the future.

Q2) Please tell us why the vehicle identification method was helpful or not helpful. In addition, let us know about the items to be supplemented or added.

# Appendix B: Additional analysis

**B1. SEM with demographic**

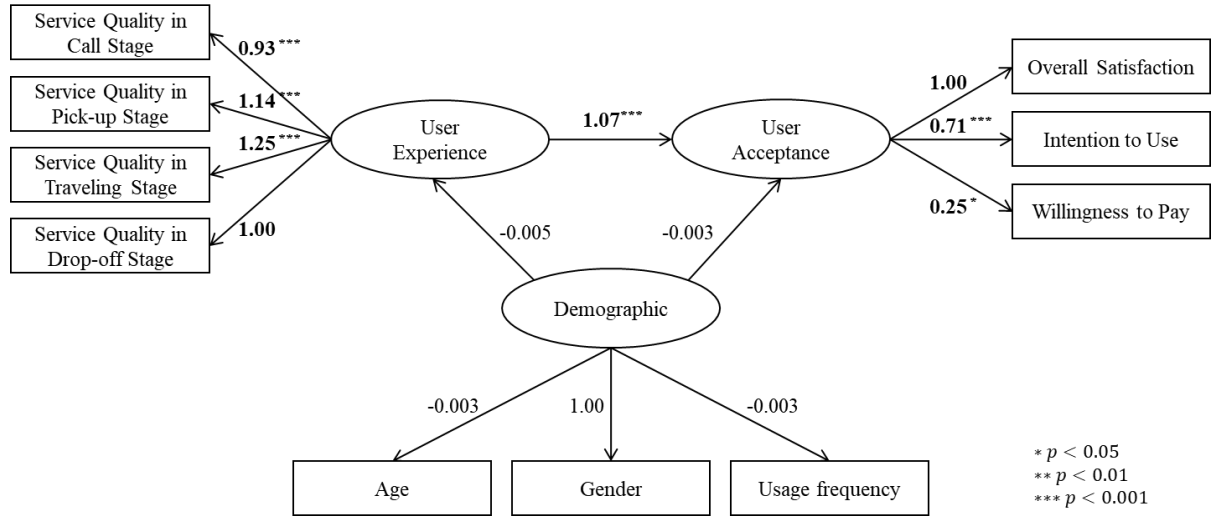

Fig B1. Relationship between user experience, user acceptance, and demographic as a Structural Equation Model

Figure B1 is the model B with demographic. The results of the goodness of fit test of the model showed that CFI = 0.84, GFI = 0.85, RMR = 0.09, RMSEA = 0.13. RMSEA deviated slightly from the threshold criterion, yet showed an acceptable fit.

In addition, we calculated the gender, age, and usage frequency either directly to the user experience or user acceptance without going through the demographic potential variable by SEM. When all three were put into the user experience, all of them had insignificant results, and when they were put into user acceptance, only age was significant. The model results are presented below Table B1. The results of goodness showed that NFI = 0.731, GFI = 0.846, RMR = 0.089, and RMSEA = 0.120.

Table B1. Standardized regression weight

|  | Estimate | P-value |
|---|---|---|
| User Acceptance ← User Experience | 0.793 | *** |
| Service Quality in Call stage ← User Experience | 0.619 | *** |
| Service Quality in Pick-up stage ← User Experience | 0.804 | *** |
| Service Quality in Traveling stage ← User Experience | 0.896 | *** |
| Service Quality in Drop-off stage ← User Experience | 0.626 |  |
| Overall Satisfaction ← User Acceptance | 0.891 |  |
| Intension to User ← User Acceptance | 0.612 | *** |
| Willingness to Pay ← User Acceptance | 0.227 | 0.079 |
| Age ← User Acceptance | -0.329 | 0.011 |
| Gender ← User Acceptance | -0.114 | 0.377 |
| Usage frequency ← User Acceptance | 0.069 | 0.595 |

## B2. Heterogeneity analysis

The results for each stage are listed in Table B2. In the pick-up, traveling, and drop-off stages excluding the call stage, the difference between females and males was significant. On average, males were more satisfied than females.

Table B2. Results of T-test between gender and each 4 stage

|  | Male | Female | p-value |
|---|---|---|---|
| Call stage | 5.50 | 5.30 | 0.478 |
| Pick-up stage | 5.96 | 5.46 | 0.044 ** |
| Traveling stage | 5.96 | 5.43 | 0.032 ** |
| Drop-off stage | 5.90 | 5.23 | 0.016 ** |

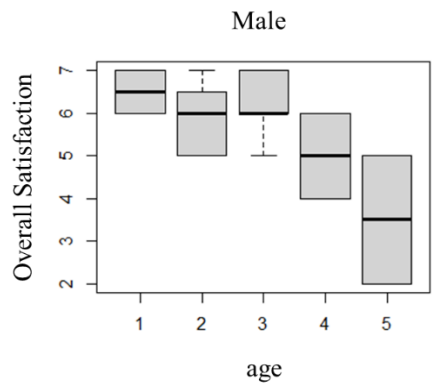

Figure B2. Box plot of Male's age and overall satisfaction

Gender differences according to age differences were analyzed using one-way ANOVA. Female overall satisfaction was similar regardless of age, but overall male satisfaction decreased significantly with age. As shown in Figure B2, 1, 2, and 3 were similar, but from 4, the satisfaction rate dropped sharply. The results of the one-way ANOVA test for male age and overall satisfaction showed a p-value of <0.01. p-value was 0.009. One is in teens, two in their 20s, three in their 30s, four in their 40s, and five in their 50s. There was a significant difference between men in their 50s, teenagers, 20s, and 30s. The p-values were 0.021, 0.013, and 0.010, respectively, and all showed significant results at a p-value <0.05. When looking at the distribution of intention to use according to the age of males and females in Figure B3, males and females showed different patterns. In males aged,> 50 years, intention to use was significantly lower, whereas, in females, the average intention to use was higher with age.

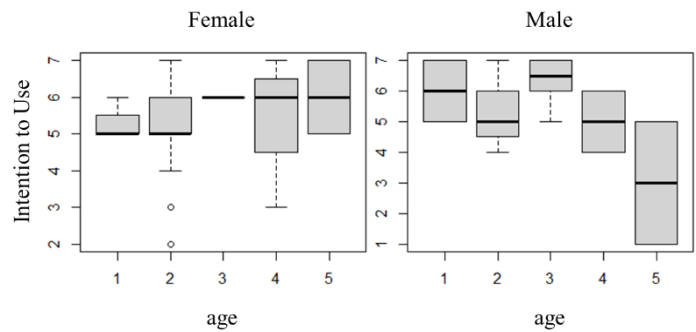

Figure B3. Box plot of age and Intention to Use (Left Female, Right Male)

**B3. Distribution of the responses**

The distribution of service quality in each stage is shown in figure B4. The stages with the highest average values are pick-up stage 5.69, traveling stage 5.68, drop-off stage 5.54, and call stage 5.39. The distributions of overall satisfaction, intention to use and WTP are shown in figure B5. The average overall satisfaction was 5.33, the average intention to use was 5.33, and the average WTP was 2.99.

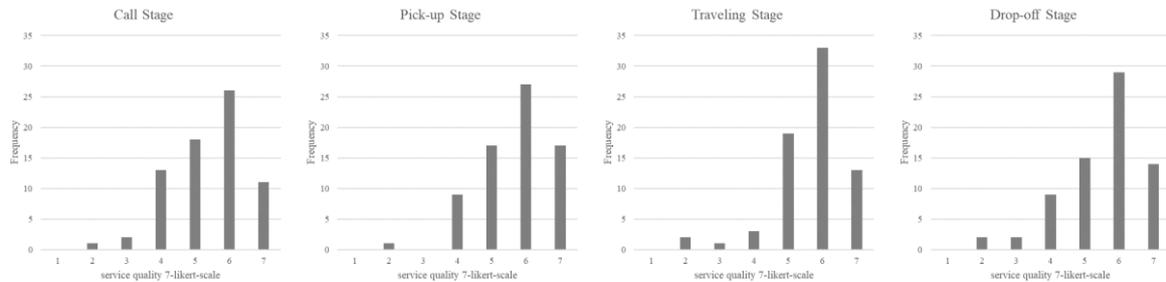

Figure B4. Distribution of the service quality in each stage

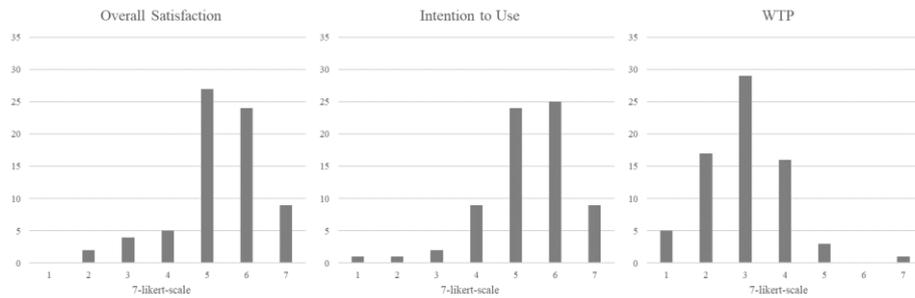

Figure B5. Distribution of Overall Satisfaction, Intention to Use, and WTP

The average service quality of the pick-up stage was the highest in the distribution of the service quality in each stage. "What were you uncomfortable with at the pick-up stage?" In response to the question, 54% of the participants answered that they were comfortable. The pick-up stage had the highest average service quality and the lowest number of complaints because the process was similar to that of using a regular taxi. The interview below is the opinion of 54% of the participants who felt comfortable about the pick-up stage.

*"It was the same as a regular taxi, so there was no inconvenience." (p7)*

*"I didn't feel any discomfort." (p14)*

*"There was no difficulty because I was provided with information on where the taxi came from." (p20)*

The analysis of the traveling and drop-off stages is covered in Sections 6.4 6.5. The call stage showed the lowest average service quality, which was judged to indicate that participants who were adapted to the latest taxi app felt the difference in skill level in the app created for experimentation. Participants made many comparisons with the current apps. Below is the content of the interviews.

*"The app is uncomfortable. It's not intuitive." (p18)*

*"Like Uber, it would be nice to know when it arrives." (p31)*

*"The app didn't have many functions." (p71)*

The distributions of overall satisfaction, intention to use and WTP were examined. The average overall satisfaction and intention to use was 5.33, which was higher than the median value. Participants showed high satisfaction and willingness to use the Robo-taxi. WTP averaged 2.99, and participants thought the reasonable fare for Robo-taxi was 75 %–99% of that of regular taxis.

**B4. Correlation between the elements at each stage**

We analyzed the correlations between the elements at each stage. The correlations between the variables of the call stage are presented in Table B3. At the call stage, the correlation between convenience and kindness, convenience and information, promptness, and reliability was high, above 0.6. The variance of the correlation of the call stage variables was the smallest among the four stages. All the variables showed similar correlations.

Table B3. Correlation between call stage variables

|                   | 1        | 2        | 3        | 4        | 5        | 6 |
|-------------------|----------|----------|----------|----------|----------|---|
| 1. Reliability    | 1        |          |          |          |          |   |
| 2. Predictability | 0.521*** | 1        |          |          |          |   |
| 3. Information    | 0.550*** | 0.586*** | 1        |          |          |   |
| 4. Kindness       | 0.514*** | 0.420**  | 0.576*** | 1        |          |   |
| 5. Convenience    | 0.455*** | 0.501*** | 0.612*** | 0.629*** | 1        |   |
| 6. Promptness     | 0.609*** | 0.390**  | 0.541*** | 0.470*** | 0.533*** | 1 |

*p<0.05, **p<0.01, ***p<0.001

The correlations between the variables of the pick-up stage are presented in Table B4. In the pick-up stage, accessibility and punctuality had the highest correlation. Participants were able to predict the pick-up process according to how well the participant communicated with the app when and where the Robo-taxi was coming, thus predictably increased reliability. Below is an interview that gave an opinion on communication during the pick-up process.

*"It was difficult to determine if it was the Robo-taxi I called." (p29)*

*"It would have been more convenient if the taxi had called the passenger directly rather than the passenger checking the number and location of the taxi." (p5)*

Table B4. Correlation between Pick-up stage variables stage

|                   | 1        | 2        | 3        | 4        | 5        | 6        | 7        | 8      | 9 |
|-------------------|----------|----------|----------|----------|----------|----------|----------|--------|---|
| 1. Reliability    | 1        |          |          |          |          |          |          |        |   |
| 2. Safety         | 0.693*** | 1        |          |          |          |          |          |        |   |
| 3. Predictability | 0.720*** | 0.583*** | 1        |          |          |          |          |        |   |
| 4. Information    | 0.498*** | 0.562*** | 0.495*** | 1        |          |          |          |        |   |
| 5. Accessibility  | 0.488*** | 0.387**  | 0.402**  | 0.477*** | 1        |          |          |        |   |
| 6. Punctuality    | 0.521*** | 0.572*** | 0.384**  | 0.563*** | 0.800*** | 1        |          |        |   |
| 7. Kindness       | 0.546*** | 0.468*** | 0.413**  | 0.588*** | 0.353**  | 0.495*** | 1        |        |   |
| 8. Communication  | 0.699*** | 0.605*** | 0.738*** | 0.552*** | 0.465*** | 0.494*** | 0.544*** | 1      |   |
| 9. Confirmation   | 0.225    | 0.297*   | 0.182    | 0.451*** | 0.402**  | 0.432*** | 0.211    | 0.259* | 1 |

*p<0.05, **p<0.01, ***p<0.001

The correlations between the variables of the traveling stage are presented in Table B5. In the traveling stage, predictability and information were correlated with reliability. Predictability and information were also correlated. The more information is given, the more participants can predict the process, and if it is predictable through the information, the reliability increases. Among the interviews, participants had the opinion that if more detailed education or guidelines were given about Robo-taxis, would be less anxious. In addition, the information was correlated with kindness. Giving the participant enough information to be satisfied makes them feel kind. Robo-taxis do not have the sensibility of regular taxis; however, we can see that they can convey the sensibility of a regular taxi through information and communication.

*"I felt that the announcement was kind." (p70)*

*"I asked about the expected arrival time, the information was provided correctly, and I was satisfied that it was the same." (p68)*

Table B5. Correlation between Traveling stage variables stage

|    | 1 | 2 | 3 | 4 | 5 | 6 | 7 | 8 | 9 | 10 | 11 |
|---|---|---|---|---|---|---|---|---|---|---|---|
| 1. Reliability | 1 | | | | | | | | | | |
| 2. Speed | 0.505*** | 1 | | | | | | | | | |
| 3. Ride comfort | 0.673*** | 0.652*** | 1 | | | | | | | | |
| 4. Safety | 0.691*** | 0.529*** | 0.669*** | 1 | | | | | | | |
| 5. Predictability | **0.740*** ** | 0.442*** | 0.529*** | 0.472*** | 1 | | | | | | |
| 6. Information | **0.716*** ** | 0.415** | 0.546*** | 0.547*** | **0.760*** ** | 1 | | | | | |
| 7. Kindness | 0.631*** | 0.342** | 0.574*** | 0.571*** | 0.611*** | **0.701*** ** | 1 | | | | |
| 8. Communication | 0.629*** | 0.456*** | 0.548*** | 0.646*** | 0.576*** | 0.685*** | 0.685*** | 1 | | | |
| 9. Pleasantness | 0.581*** | 0.349** | 0.583*** | 0.480*** | 0.447*** | 0.457*** | 0.677*** | 0.483*** | 1 | | |
| 10. Convenience | 0.616*** | 0.416** | 0.460*** | 0.556*** | 0.385** | 0.463*** | 0.511*** | 0.677*** | 0.512*** | 1 | |
| 11. Comfort | 0.359** | 0.413** | 0.340** | 0.481*** | 0.296 | 0.334** | 0.364** | 0.299* | 0.441*** | 0.404** | 1 |

$*p<0.05, **p<0.01, ***p<0.001$

The correlations between the variables of the drop-off stage are presented in Table B6. In the drop-off stage, communication and kindness had the greatest correlation. Communication with Robo-taxis was correlated with the kind of feelings the participants received. Reliability, safety, and predictability were correlated, and predictability was correlated with information. The relationship between reliability, predictability, and information was similar in other stages; on the other hand, the difference in the drop-off stage was that safety was correlated with reliability. At the drop-off stage, participants judged that getting off safely and predictably was important for reliability. When getting off a taxi, the conditions on the road were different from time to time, so the participants made it important to get off safely.

*"I felt anxious when stopping because there was a vehicle in front of taxi." (p11)*

*"The car was crowded near the destination. I was worried about where the Robo-taxi would drop off safely." (p65)*

Table B6. Correlation between Drop-off stage variables stage

|    | 1 | 2 | 3 | 4 | 5 | 6 | 7 | 8 |
|---|---|---|---|---|---|---|---|---|
| 1. Reliability | 1 | | | | | | | |
| 2. Safety | **0.742*** ** | 1 | | | | | | |
| 3. Predictability | **0.737*** ** | 0.527*** | 1 | | | | | |
| 4. Information | 0.680*** | 0.594*** | **0.789*** ** | 1 | | | | |
| 5. Accessibility | 0.438*** | 0.299* | 0.447*** | 0.320* | 1 | | | |
| 6. Punctuality | 0.613*** | 0.485*** | 0.486*** | 0.462*** | 0.699*** | 1 | | |
| 7. Communication | 0.604*** | 0.615*** | 0.640*** | 0.646*** | 0.440*** | 0.572*** | 1 | |
| 8. Kindness | 0.576*** | 0.615*** | 0.571*** | 0.631*** | 0.287* | 0.452*** | **0.828*** ** | 1 |

$*p<0.05, **p<0.01, ***p<0.001$

As a result of examining the correlation of each stage, the correlation with reliability was important. The higher the predictability, the better the reliability at all stages. In the pick-up stage, communication was important for finding a taxi. In the traveling stage, information was important, and information was correlated with kindness. The more information about the traveling stage, the more the participants felt the kindness and reliability of the Robo-taxi. Safety is important in the drop-off stage. The more you can get off safely, the greater is the reliability. In the drop-off stage, there was a high correlation between communication and kindness. Participants felt kindness in communication with Robo-taxis to get off safely.

**B5. PCA has two factors**

Table B7. Results of PCA of 24 emotional factors

| Factor | Component | |
|---|---|---|
| | 1 | 2 |
| **Trendy** | **0.885** | |
| **Ingenious** | **0.834** | |
| **Sophisticated** | **0.825** | |
| New | 0.792 | |
| Simple | 0.768 | |
| Excellent | 0.745 | |
| Convenient | 0.742 | |
| Efficient | 0.711 | |
| Disappointing | -0.668 | |
| Frustrating | -0.637 | |
| Annoying, | -0.576 | |
| Strange | -0.560 | |
| Complicated | -0.549 | |
| Unpleasant | -0.527 | |
| Tiresome | -0.510 | |
| Stuffy | -0.451 | |
| Familiar | 0.434 | |
| Dull | -0.307 | |
| **Uncomfortable** | | **0.885** |
| **Afraid** | | **0.862** |
| **Nervous** | | **0.884** |
| Safe | | -0.755 |
| Reliable | | -0.579 |
| Comfortable | | -0.575 |

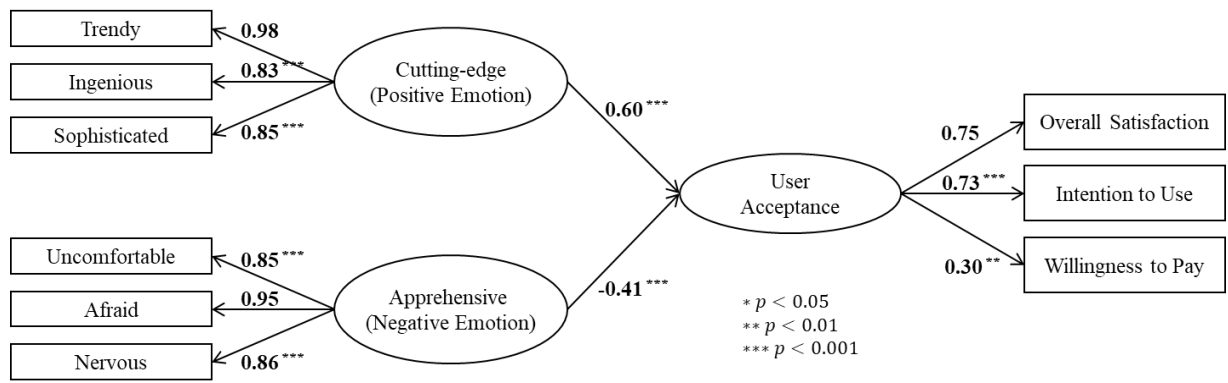

Figure B6. SEM showing the relationship between positive and negative emotions and user acceptance (Model C)

The total cumulative explanatory amount of PCA was 55.285%. The results of the goodness of fit test showed that NFI = 0.95, GFI = 0.95, RMR = 0.10, and RMSEA = 0.01, thus showing acceptable fitness.

**B6. PCA has four factors**

Table B8. The results of PCA of 24 emotional factors

| Factor | Component | | | |
|---|---|---|---|---|
| | 1 | 2 | 3 | 4 |
| **New** | **0.903** | | | |
| **Ingenious** | **0.885** | | | |
| **Trendy** | **0.857** | | | |
| Excellent | 0.768 | | | |
| Sophisticated | 0.726 | | | |
| Efficient | 0.699 | | | |
| Simple | 0.688 | | | |
| Convenient | 0.601 | | | |
| Familiar | 0.573 | | | |
| Frustrating | -0.466 | | | |
| **Unpleasant** | | **0.850** | | |
| **Annoying** | | **0.816** | | |
| **Tiresome** | | **0.746** | | |
| Strange | | 0.709 | | |
| Complicated | | 0.693 | | |
| Disappointed | | 0.693 | | |
| **Uncomfortable** | | | **0.882** | |
| **Nervous** | | | **0.840** | |
| **Afraid** | | | **0.826** | |
| Safe | | | -0.775 | |
| Reliable | | | -0.595 | |
| Comfortable | | | -0.587 | |
| **Stuffy** | | | | **0.789** |
| **Dull** | | | | **0.681** |

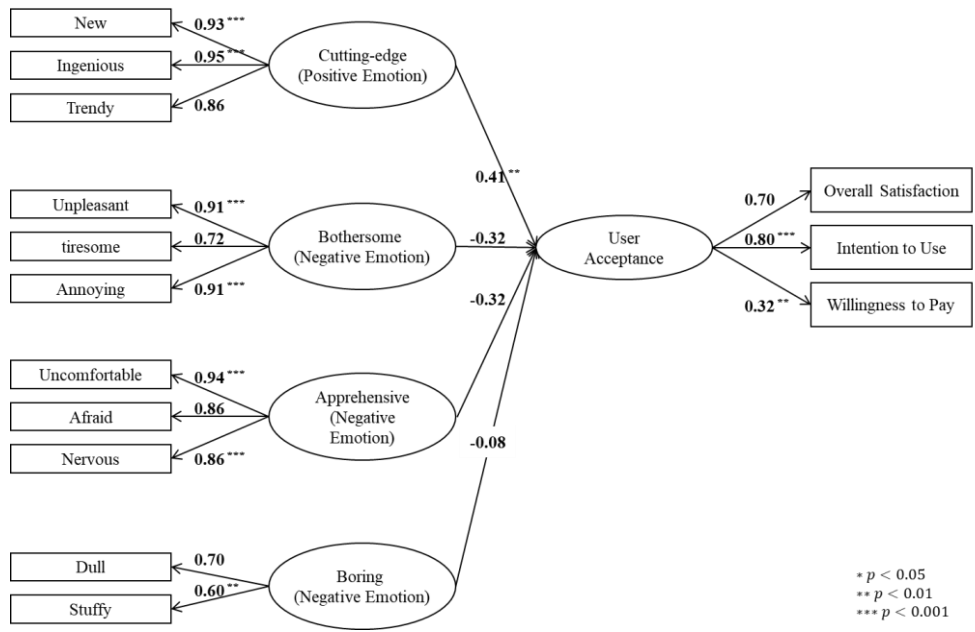

Figure B7. SEM showing the relationship between emotions and user acceptance (Model C)

The total cumulative explanatory amount of PCA was 71.104%. The results of the goodness of fit test showed that NFI = 0.88, GFI = 0.88, RMR = 0.12, and RMSEA = 0.05, thus showing acceptable fitness.

## B7. Algebraic definition of model fit

Table B9. Algebraic definition of model fit (64)

| Algebraic definition | Property |
|---|---|
| $\text{NFI} = \dfrac{X^2(\text{Null model}) - X^2(\text{Proposed model})}{X^2(\text{Null model})}$ | NFI ≥ 0.8-0.9 |
| $\text{GFI} = 1 - \left[\dfrac{tr(\Sigma^{-1}\mathbf{S}-1)^2}{tr(\Sigma^{-1}\mathbf{S})^2}\right]$ | GFI ≥ 0.8-0.9 |
| $\text{RMR} = \sqrt{\dfrac{\left\{2\sum_{i=1}^{P}\sum_{j=1}^{i}\left[\dfrac{s_{ij}-\hat{\sigma}_{ij}}{s_{ii}s_{jj}}\right]^2\right\}}{p(p+1)}}$ | RMR ≤ 0.08 |
| $\text{RMSEA} = \sqrt{\dfrac{X^2 - df}{df(N-1)}}$ | RMSEA ≤ 0.08 |

$X^2$ is the chi-square value for each model. *tr* is trace of a matrix. **S** is the sample covariance matrix. $s_{ij}$ is observed covariance. $\hat{\sigma}_{ij}$ is reproduced covariance. $p$ is the number of measured variables. $df$ is the degrees of freedom. N is the sample size.